\documentclass[10pt, conference]{IEEEtran}
\IEEEoverridecommandlockouts
\usepackage{cite}
\usepackage{amsmath,amssymb,amsfonts}
\usepackage{algorithmic}
\usepackage{listings}
\usepackage{graphicx}
\usepackage{array}
\usepackage{textcomp}
\usepackage{xcolor}
\usepackage{subcaption}
\usepackage{booktabs}
\def\BibTeX{{\rm B\kern-.05em{\sc i\kern-.025em b}\kern-.08em
    T\kern-.1667em\lower.7ex\hbox{E}\kern-.125emX}}
\begin{document}


\definecolor{forestgreen}{RGB}{34,139,34}
\definecolor{orangered}{RGB}{239,134,64}
\definecolor{darkblue}{rgb}{0.0,0.0,0.6}
\definecolor{gray}{rgb}{0.4,0.4,0.4}
\definecolor{codegreen}{rgb}{0,0.6,0}
\definecolor{codegray}{rgb}{0.5,0.5,0.5}
\definecolor{codepurple}{rgb}{0.58,0,0.82}
\definecolor{backcolour}{rgb}{0.95,0.95,0.92}

\lstdefinestyle{XML} {
    language=XML,
    extendedchars=true, 
    breaklines=true,
    breakatwhitespace=true,
    emph={},
    emphstyle=\color{red},
    basicstyle=\ttfamily,
    columns=fullflexible,
    commentstyle=\color{gray}\upshape,
    morestring=[b]",
    morecomment=[s]{<?}{?>},
    morecomment=[s][\color{forestgreen}]{<!--}{-->},
    keywordstyle=\color{orangered},
    stringstyle=\ttfamily\color{black}\normalfont,
    tagstyle=\color{darkblue}\bf,
    morekeywords={attribute,xmlns,version,type,release},
    otherkeywords={attribute=, xmlns=},
}

\lstdefinelanguage{PDDL}
{
  sensitive=false,    
  morecomment=[l]{;}, 
  alsoletter={:,-},   
  morekeywords={
    define,domain,problem,not,and,or,when,forall,exists,either,
    :domain,:requirements,:types,:objects,:constants,
    :predicates,:action,:parameters,:precondition,:effect,
    :fluents,:primary-effect,:side-effect,:init,:goal,
    :strips,:adl,:equality,:typing,:conditional-effects,
    :negative-preconditions,:disjunctive-preconditions,
    :existential-preconditions,:universal-preconditions,:quantified-preconditions,
    :functions,assign,increase,decrease,scale-up,scale-down,
    :metric,minimize,maximize,
    :durative-actions,:duration-inequalities,:continuous-effects,
    :durative-action,:duration,:condition
  }
}

\lstdefinelanguage{Srv}{
keywords = [1]{bool, uint8, int32, uint64, float32, float64, string, Header, Point, Quaternion, time},
comment=[l]{\#}
}

\lstdefinestyle{mystyle}{
  backgroundcolor=\color{backcolour},   commentstyle=\color{codegray},
  keywordstyle=\color{codegreen},
  numberstyle=\tiny\color{codegray},
  stringstyle=\color{codepurple},
  basicstyle=\ttfamily\footnotesize,
  breakatwhitespace=false,         
  breaklines=true,                 
  captionpos=b,                    
  keepspaces=true,                 
  numbers=left,                    
  numbersep=5pt,                  
  showspaces=false,                
  showstringspaces=false,
  showtabs=false,                  
  tabsize=2
}
\lstset{style=mystyle}

\title{Enhancing Human-in-the-Loop Adaptive Systems through Digital Twins and VR Interfaces
}

\author{Enes Yigitbas, Kadiray Karakaya, Ivan Jovanovikj, Gregor Engels  \\
\textit{Department of Computer Science} \\
\textit{Paderborn University}\\
Paderborn, Germany \\
firstname.lastname@uni-paderborn.de}

\maketitle

\begin{abstract}
Self-adaptation approaches usually rely on closed-loop controllers that avoid human intervention from adaptation. While such fully automated approaches have proven successful in many application domains, there are situations where human involvement in the adaptation process is beneficial or even necessary. For such "human-in-the-loop" adaptive systems, two major challenges, namely transparency and controllability, have to be addressed to include the human in the self-adaptation loop. Transparency means that relevant context information about the adaptive systems and its context is represented based on a digital twin enabling the human an immersive and realistic view. Concerning controllability, the decision-making and adaptation operations should be managed in a natural and interactive way. As existing human-in-the-loop adaptation approaches do not fully cover these aspects, we investigate alternative human-in-the-loop strategies by using a combination of digital twins and virtual reality (VR) interfaces. Based on the concept of the digital twin, we represent a self-adaptive system and its respective context in a virtual environment. With the help of a VR interface, we support an immersive and realistic human involvement in the self-adaptation loop by mirroring the physical entities of the real world to the VR interface. For integrating the human in the decision-making and adaptation process, we have implemented and analyzed two different human-in-the-loop strategies in VR: a procedural control where the human can control the decision making-process and adaptations through VR interactions (human-controlled) and a declarative control where the human specifies the goal state and the configuration is delegated to an AI planner (mixed-initiative). We illustrate and evaluate our approach based on an autonomic robot system that is accessible and controlled through a VR interface. 
\end{abstract}


\section{Introduction}
\label{intro}
Over the last decade, self-adaptive systems (SAS) have emerged as a solution to overcome many of the limitations of human supervision by endowing systems with mechanisms to automatically adapt their structure and behavior at run time~\cite{seams15_human_participation}. Typical application areas of SAS are for example robotics~\cite{edwards2009architecture}, control systems~\cite{filieri2015software}, software architectures~\cite{DBLP:conf/models/NagelGYCE12}, fault-tolerant computing~\cite{ebnenasir2007designing}, or smart user interfaces~\cite{DBLP:conf/ecmdafa/YigitbasS0E17}. While most of the existing self-adaptation approaches rely on closed-loop controllers (e.g., implementing the MAPE-K reference framework) that eliminate human intervention from adaptation, there are application domains where it is beneficial to integrate the human in the MAPE-K loop~\cite{seams15_human_participation}. One of these application domains are autonomous robots which have a great potential to help for example in disaster scenarios. Although autonomous robots can operate in human unfriendly or hazardous environments, designing software for such self-adaptive systems is difficult as the robots need to deal with unknown situations that cannot be completely anticipated at design time~\cite{DBLP:conf/icse/NiemczykG15}. In such a scenario with uncertain circumstances, the different activities of the MAPE-K loop in some classes of systems (e.g., safety-critical) and application domains can benefit from human involvement~\cite{seams15_human_participation}. However, the integration of the human in the MAPE-K loop brings various challenges that should be addressed both from self-adaptive systems and human-computer interaction perspective.

Challenge 1: \textit{Transparency} \\
The first important challenge in incorporating the human in the MAPE-K loop is supporting transparency. This means that humans need to understand what is going on in the autonomous system as well as its context. Providing the user with feedback, explanation, and visualization about the system state and its context is crucial to establish and keep the user's trust and spatial awareness~\cite{seams16_designing_human_itl}. However, representing real-world aspects including the self-adaptive system and its context is quite complex, and relevant aspects should be mirrored to enable the user transparency about the underlying system.  

Challenge 2: \textit{Controllability} \\
The second important challenge for human involvement in the MAPE-K loop is managing the degree and way of controllability. Concerning the degree of controllability, to ensure that the most suitable decisions are met, it is important that self-adaptive systems provide mechanisms to adjust the degree of user control (e.g., fully human-controlled or mixed-initiative) depending on different circumstances (e.g., changing goals, emergent behaviors, uncertainties, or simply that users prefer more control). Regarding the way of controllability, incorporating human input in the decision-making and adaptation process should be supported in a natural and interactive way.

To address the above-mentioned challenges, we investigate how human-in-the-loop adaptive systems can be enhanced by using a combination of digital twins~\cite{DBLP:conf/icse/JosifovskaYE19} and virtual reality interfaces~\cite{PAN200620}. For supporting transparency, according to~\cite{DBLP:conf/iot/KritzlerFMR17}, we create a digital-twin based virtual representation of a self-adaptive system and its context where the real-world setting is realized in a virtual environment. The virtual environment can be accessed and controlled by a human with the help of a virtual reality (VR) interface. This VR interface supports observing the digital twin (3D model) by mirroring the physical entities of the real world including the SAS and its context. Besides that, it supports controllability based on two different types of human-in-the-loop control. The first type of control, called procedural control, supports the integration of the human by enabling the end-user to control the decision-making and adaptation process through virtual reality interactions (e.g., grabbing, moving a robot arm). The sequence of VR interactions can be recorded and sent as a command to the robot enabling unique adaptations and even the addition of new operational sequences. While this kind of human-in-the-loop integration is fully controlled by the human, an alternative way of incorporating human input in the decision making and adaptation process would rely on the mixed-initiative approach. For reaching this, we implemented a second type of control in the VR interface, called declarative control, where the end-user can specify the goal states for a self-adaptive system based on VR interactions and the behavior of the system is computed through an AI planner. For analyzing and comparing the advantages and disadvantages of both human-in-the-loop solution variants, we have evaluated each solution regarding efficiency, effectiveness, and user satisfaction. 

This paper is organized as follows. In Section \ref{scenario}, we present Dobot Magician, the robot system that we employ to illustrate our approach. Section \ref{rel_work} introduces related work in the areas of self-adaptive systems, autonomous robots, and virtual reality interfaces, outlining some of the major requirements for the solution approach. Section \ref{sol_idea} details the conceptual solution idea and Section \ref{impl} describes its implementation. In Section \ref{eval}, we present the evaluation setup and its main results. Finally, Section \ref{sum} presents conclusions and directions for future work. 

\section{Motivating Scenario}
\label{scenario}
For motivating our approach, we employ an industrial assembly scenario. In this scenario, a robot system, which operates on an assembly line, consists of a Kinect camera to detect the objects in the environment and a robot arm to move the objects. The task of the robot in such a scenario is to put various objects together to build something. However, the positions of the objects or the obstacles in the environment can not be assumed unchanging. Therefore, the robot needs to adapt to such changes in its environment. The adaptation corresponds to changing the robot arm's motion trajectory. Depending on the safety criticality of the operation domain, varying degrees of human supervision can be required to intervene in these motions, e.g., some assembly tasks may require more precise and sensitive movements, while some tasks can benefit from more autonomy.

\begin{figure}[hbt!]
\centering
\includegraphics[width=252pt,keepaspectratio]{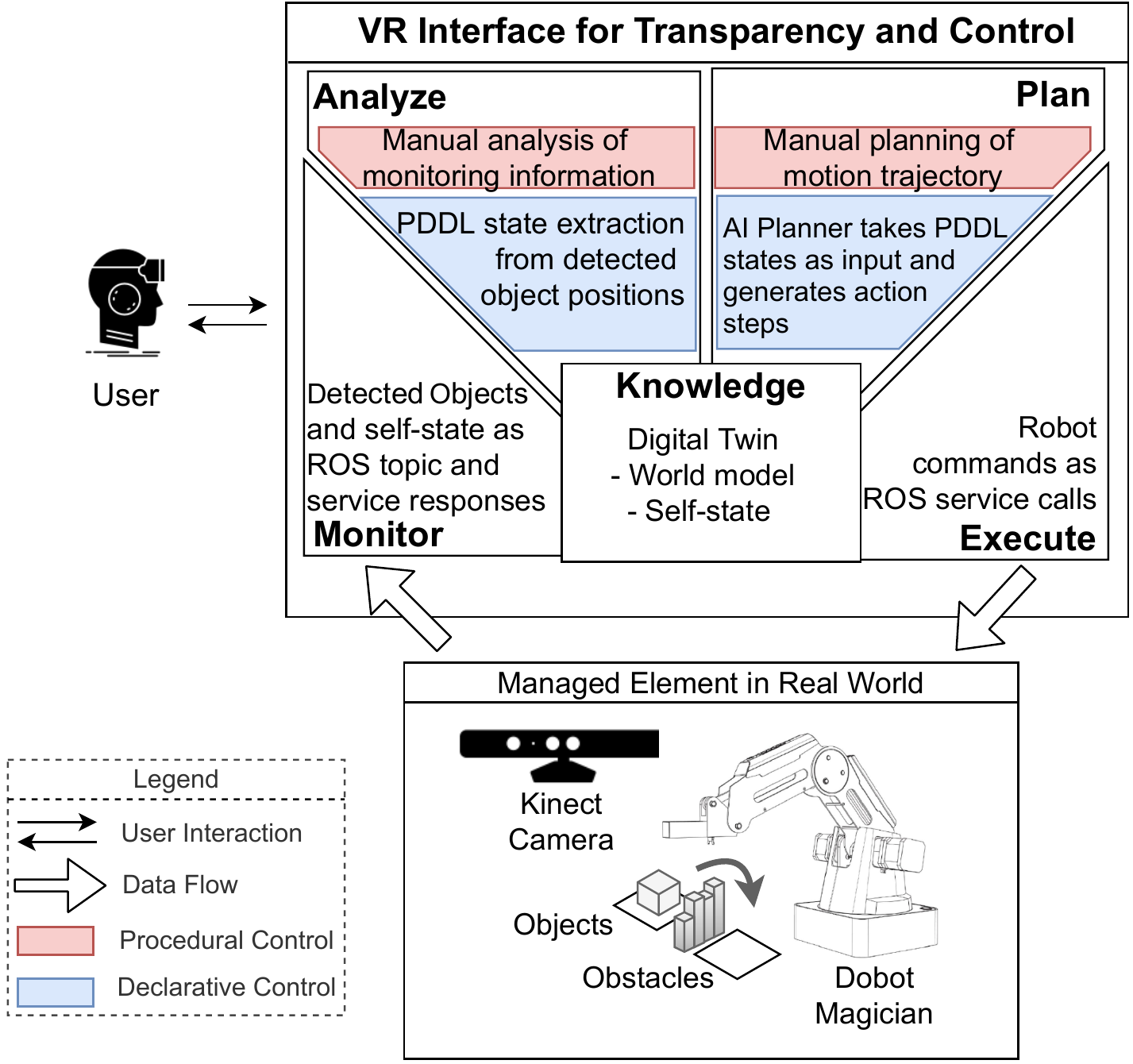}
\caption{Motivating scenario: supporting human involvement in the MAPE-K loop through a virtual reality interface}
\label{fig:scenario}
\end{figure}

Fig. \ref{fig:scenario}, shows the overview of our motivating scenario, where the Dobot Magician system that runs on ROS (Robot Operating System) is the \textit{Managed Element}. The system is monitored through ROS topics that publish visual data obtained by the Kinect camera, and through ROS services that return the robot's self-state. The \textit{monitoring} information is used to reflect the digital twin of the system in a VR application, for providing transparency to the user about the system state. To provide varying degrees of controllability, the VR interface offers a procedural control and a declarative control. With the procedural control, the user intervenes in the loop by manually \textit{analyzing} the visualized \textit{monitoring} information. After that, the user manually \textit{plans} the exact motion trajectory that the robot arm should follow. Eventually, the user sends the defined motions as robot commands for \textit{execution} by the real robot. With the declarative control, the information about object positions is \textit{analyzed} to extract the initial state in PDDL (Planning Domain Description Language) \cite{pddl_1998}. The user intervenes in the loop to define the \textit{task} by creating the goal state, a combination of object positions that represent a target assembly. The extracted initial state and user-defined goal state are then combined to create a PDDL problem and sent to an AI planner, which returns a set of actions as the \textit{plan} that leads from the initial state to the goal state. To visualize the plan to the user for transparency, it is first executed by the digital twin of the robot. If the plan is satisfactory, the user approves it to be \textit{executed} by the real robot.

\section{Related Work}
\label{rel_work}
Recent approaches in the area of self-adaptive systems (SAS) already addressed the need for human-in-the-loop systems. Therefore, in this section, we begin with a summary of the developments in the field regarding human-in-the-loop self-adaptive systems. Conforming to our applied scenario, we continue with the relevant approaches in the field of human-robot interaction (HRI) and conclude with an overview of recent VR interfaces for HRI.

Although self-adaptive systems are designed to be self-managing without human intervention, they are shown to benefit from human involvement. \cite{brun2009engineering} argues that self-adaptive systems need to keep humans in the loop to gain the trust of the user by providing feedback (transparency) about the system state. \cite{seams16_designing_human_itl} presents a set of requirements that SAS should support including transparency and controllability, and presents the stages of the adaptation loop that humans can participate in. It suggests, that humans can involve in the loop by taking the roles of sensor, actuator, decision-maker, or knowledge augmenter. Following this, various approaches aim to take advantage of human involvement in adaptive systems. \cite{whittle_role_2010} demonstrates the benefits of involving humans in the monitoring phase, to provide input in cases where complete monitoring of the environment is not possible. \cite{evers_user_2014} reports, that proving control to the users is preferred, by assessing several types of human participation (implicit, explicit) in different phases of the MAPE-K loop. \cite{seams15_human_participation} presents a formal framework, where a human participates in the execute stage and shows that it can improve system utility. \cite{seams17_improving_hitl_brain} shows that the mental state of the human in the loop can affect the system performance, and whether to involve human at all should also be considered. Despite the repeatedly shown benefits, finding an optimal degree of human intervention that maximizes system performance remains to be an open question. 

To make a step towards this direction, we investigate the benefits of different degrees of human involvement in an adaptive robot system.
Since our concrete implementation is based on a robot system, we consider related work in the HRI field. \cite{hri_survey} presents a scale for the level of autonomy in HRI. On one edge of the scale is teleoperation, e.g.,~\cite{robot_teleop_vr_2002},\cite{robot_teleop_space_2004},\cite{robot_teleop_surgery_2004}, where the human has the full control, and on the other edge is autonomy, e.g.,~\cite{robot_autonom_1995},\cite{robot_autonom_2010},\cite{robot_autonom_2011},\cite{robot_autonom_2019}, where the robot has the full control. \cite{lyons2013being} presents transparency as a crucial mechanism for the effective HRI, as we identified as the first challenge to address. \cite{kim_who_2006} suggests that autonomous robots should provide transparency to be more understandable to humans, but notes that the language to convey transparency must be understandable by the user. As the users of the system would have diverse technical backgrounds, the means of transparency must target inclusiveness. \cite{theodorou_designing_2017} argues that graphical representation of an autonomous system's plans is easy to follow by less-technical users. Motivated by this argumentation, to establish transparency, we employ a 3D graphical visualization of the robot, its environment, and its planned actions through a VR interface. 

In previous work, the concept of digital twin has been used in several domains such as manufacturing~\cite{tao2018digital} or assistance systems~\cite{DBLP:conf/hci/JosifovskaYE19}. Also, VR interfaces have been used in different application domains such as training~\cite{DBLP:conf/vrst/YigitbasJSE20}, education~\cite{DBLP:conf/mc/YigitbasTE20}, or healthcare~\cite{DBLP:conf/mc/YigitbasHE19}. Utilizing a combination of the concept of digital twin and VR interfaces has been shown successfully by previous works especially for HRI. A key requirement for the effectiveness of these interfaces is the provided controllability, as we identified as the second challenge to address. 
\cite{ros_reality_2018} suggests that VR interfaces can allow non-expert users to control robots, and presents a controlling mechanism where the user's hands are directly mapped to the robot's hands. To support controllability, \cite{baxters_homunculus_2018} presents a VR interface that simulates a control room (VRCR), where the user sees the world from the robot’s eyes and uses control orbs to move its arms. Another approach for improved controllability was presented by \cite{robot_prog_digitaltwin_2020}, where a VR interface including the digital twin of a robotic station, tracks user interaction with a real object for programming the robot. Similarly, \cite{varm_2017} enables controlling a robot in VR, by allowing the user to define the trajectory that the robot arm should follow by detecting hand gestures of the user.

In summary, the existing approaches serve as an initial inspiration. However, none of them fully addresses the challenges concerning transparency and controllability.

\section{Solution Approach}
\label{sol_idea}
In this section, we explain how our solution provides transparency into the system state. Subsequently, we present two different control strategies, procedural control, and declarative control, to enable different degrees of controllability for the users.

\begin{figure}[hbt!]
\centering
\includegraphics[width=252pt,keepaspectratio]{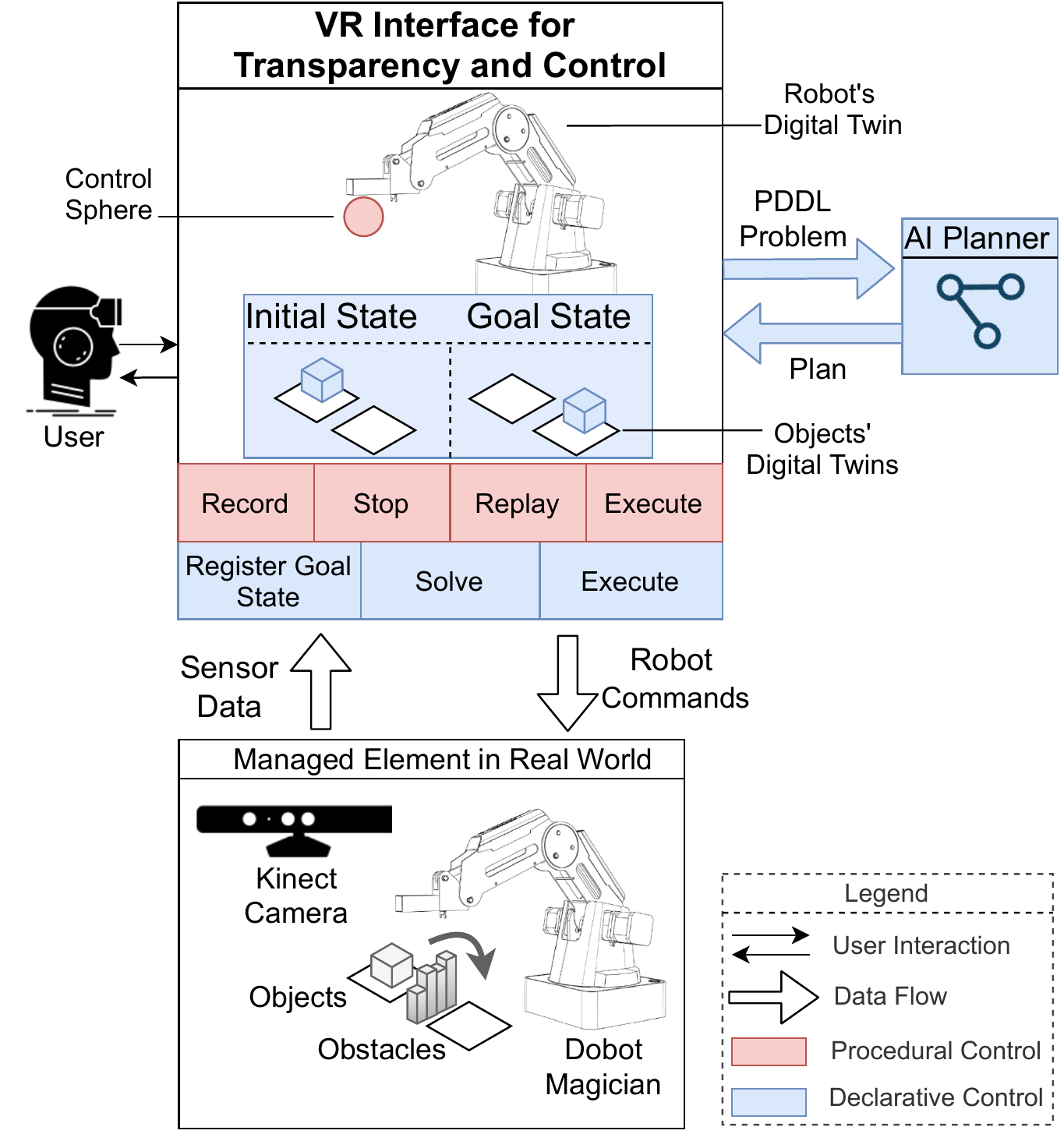}
\caption{Overview of the solution approach}
\label{fig:overview}
\end{figure}

To present the system state to the user, as depicted in Fig. \ref{fig:overview}, we have developed a \textit{VR Interface} that enables the user to observe the system in the 3D VR environment. In the presented scenario, the robot system in the real world consists of a Dobot Magician arm, and cube objects that the robot arm can pick up with its suction cup. Therefore, we create a digital twin of the robot system, as well as the digital twins of the cube objects in the VR scene. The user can walk freely in this environment, and observe the robot and objects in the environment from different viewpoints. This is the advantage of the VR interfaces compared to traditional GUIs, where the user feels fully immersed in the visualized environment. 

We visualize the system state by using two sources of input. The first source of input is the state of the robot itself. The state of the robot is managed by a ROS node and made available as a ROS service. The VR interface calls this service to obtain the position of the robot's end effector (as 3D coordinates), and its suction state (can be enabled or disabled). According to this information, the pose of the robot's digital twin is defined. The second source of input is the state of the environment, which corresponds to the positions of the objects in the environment. The objects must be precisely detected, and their digital twins must be precisely represented, e.g., the objects must appear in the VR scene according to their actual positions in the real world. For this purpose, we use a Kinect camera that is managed by a machine learning-enabled ROS node, which is trained to detect the positions and the colors of cube objects. The object detection message is streamed continuously by a ROS topic. The VR interface subscribes to this topic to obtain information about the objects. The digital twins of the cubes are created according to their positions, and their colors. For accounting for the cases, where the environment contains objects unknown to the Kinect ROS node, we present a screen in the VR scene that shows a video stream of the environment.

The first control strategy that we implemented is called procedural control. With procedural control (highlighted in red in Fig. \ref{fig:overview}), an  adaptation of the programming  by  demonstration  (PbD) approach [23]  to VR has been applied, where the user can directly  interact  with  the  robot  arm. The user can grab the end effector (control sphere in Fig. \ref{fig:overview}) of the arm to move it in the 3D space, and joints of the arm will be aligned accordingly. This is enabled by using the inverse kinematics calculation for the Dobot Magician arm \cite{dobot_ik_2017}. The user records the robot arm movements, which make up the exact motion trajectory for its end effector, and sends it to the real robot for execution. The interaction for the recording process is done via UI buttons that are touchable in the VR scene.

The second control strategy that we implemented is declarative control (highlighted in blue in Fig. \ref{fig:overview}), a mixed-initiative approach, where the user only defines a goal state by moving cube objects. In declarative control, the user can grab the cube objects to define a goal state, and the robot figures out how to reach the goal state autonomously. To enable this, the system state, which corresponds to the configuration of the cube objects, needs to be extracted. For this purpose, each object's relation to other objects is calculated and represented as PDDL predicates.

\begin{figure}[hbt!]
\centering
\begin{subfigure}{.23\textwidth}
  \centering
  \includegraphics[width=.35\linewidth]{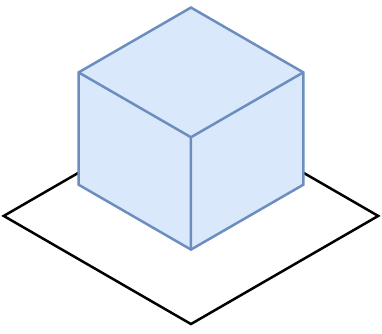}
  \caption{(at cubeA posA)}
  \label{fig:cube_at_pos}
\end{subfigure}
\begin{subfigure}{.23\textwidth}
  \centering
  \includegraphics[width=.65\linewidth]{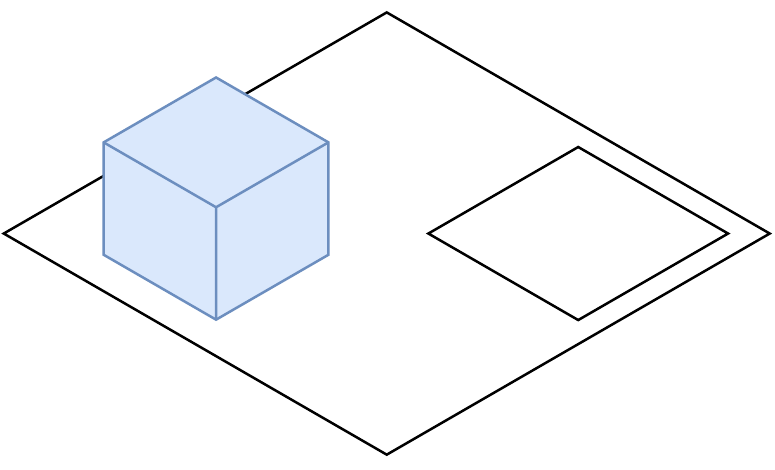}
  \caption{(ontable cubeA)}
  \label{fig:cube_on_table}
\end{subfigure}
\begin{subfigure}{.23\textwidth}
  \centering
  \includegraphics[width=.2\linewidth]{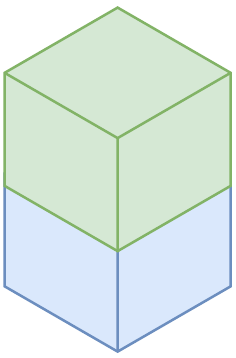}
  \caption{(on cubeB cubeA)}
  \label{fig:cube_on_cube}
\end{subfigure}
\caption{Possible cube configurations and corresponding PDDL statements}
\label{fig:cube_pddl}
\end{figure}

Fig. \ref{fig:cube_pddl}, shows some possible cube configurations and the corresponding PDDL statements. In addition to this, we also represent cubes that do not have another cube on top (e.g., (clear cubeA)), and positions that do not contain cubes (e.g., (free posA)). When the human intervention starts, the positions of the objects are registered as the initial state. Once the user is done with positioning the cube objects in the VR interface, the positions of the objects are registered as the goal state. An initial state and a goal state in PDDL makes up a PDDL problem, which can be solved by an AI planner (PDDL solver). The solution is a set of actions (plan) that leads from the initial state to the goal state. The plan is first simulated by the digital twin of the robot and presented for human approval. If the user is satisfied with the movements in the VR interface, these can be sent to the real robot for execution.

\section{Implementation}
\label{impl}

\begin{figure}[hbt!]
\centering
\begin{subfigure}{.45\textwidth}
  \centering
  \includegraphics[width=\linewidth]{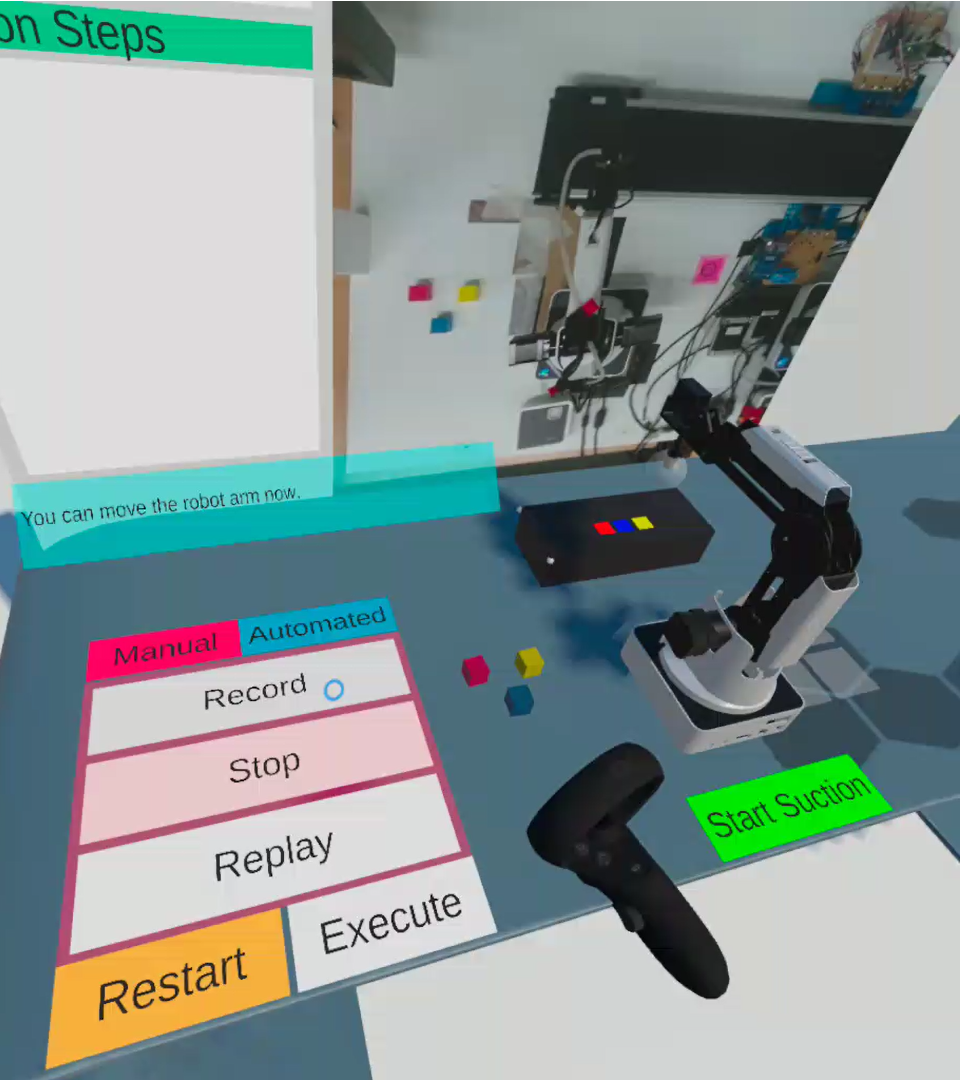}
  \caption{VR interface in procedural control}
  \label{fig:vr_proc}
\end{subfigure}
\begin{subfigure}{.45\textwidth}
  \centering
  \includegraphics[width=\linewidth]{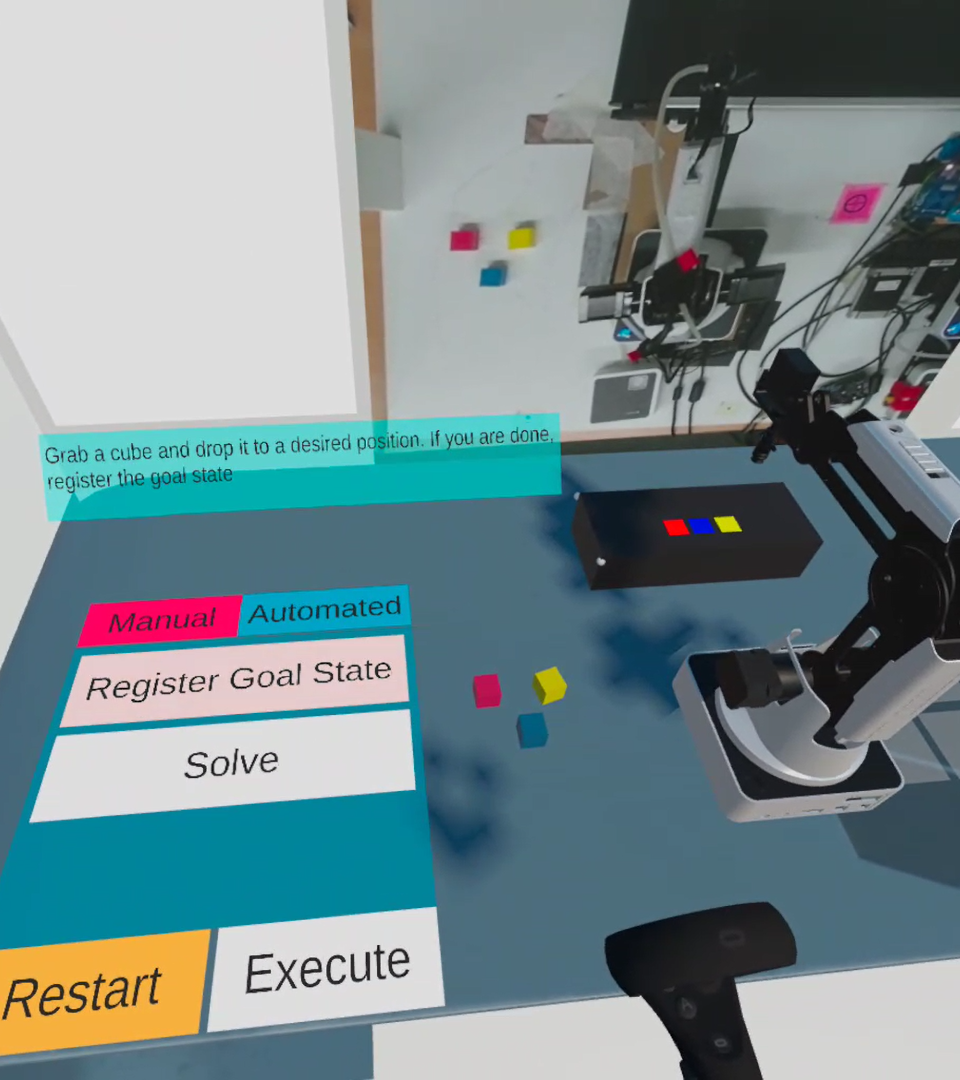}
  \caption{VR interface in declarative control}
  \label{fig:vr_dec}
\end{subfigure}
\caption{VR interface in two control strategies}
\label{fig:vr_procedural_vs_declarative}
\end{figure}

We implemented the VR interface using the Oculus Quest \cite{oculus_quest} VR headset. As with most commercial VR headsets, it consists of a Head-Mounted Display (HMD) for 3D vision, and hand controllers for input. For the implementation, we used the Unity \cite{unity_website} game engine, with the Oculus Integration \cite{oculus_integration} plugin, as a main development platform. For representing the digital twin of the Dobot Magician arm, we created its 3D model and imported it to Unity. Similarly, 3D cube models were used for creating digital twins of the cube objects. For obtaining the states of the robot and the cubes, the VR interface needs to communicate with the ROS nodes that manage the robot system. The communication was implemented using the ROS\# Unity plugin \cite{rossharp} that supports Rosbridge Protocol \cite{rosbridge}. There are two methods for communication in ROS, as ROS services and ROS topics. ROS services return a response when a client calls a service, and ROS topics stream a continuous message when a client subscribes to a topic. To support transparency, the VR interface calls a ROS service that returns the state of the robot and subscribes to two ROS topics. The first topic publishes the positions and colors of the detected objects, and the second topic publishes a video stream of the environment. As shown in Fig. \ref{fig:vr_procedural_vs_declarative}, the detected objects are instantiated with the position and color information obtained from the ROS topic. Similarly, a video stream is projected to the VR field of view which shows the top-down view of the robot system. Additionally, we model and track user interactions as a state machine, to guide the user with hints (panel in transparent blue).

Fig. \ref{fig:vr_proc} shows the VR interface when the procedural control is activated. With this control, our goal is to enable the user to directly interact with the robot arm. To give the feeling of touching and moving, we provide the user with a transparent white sphere that surrounds the end effector of the robot arm. The user can hold this sphere and move it to any point in the 3D space within the reach of the arm. Once the sphere is moved, the robot arm will arrange itself to keep its end effector within the center of the sphere. The user uses the control panel on the left side of the table. By clicking on the \textit{Record} button, the user can start a recording of movements. The recordings can be stopped by clicking on the \textit{Stop} button and then replayed by clicking on the \textit{Replay} button. If the user is not satisfied with the recording or a problem occurred during the interaction,  it can be started from scratch by using the \textit{Restart} button. If the user is happy with the recorded movements, the recorded movements can be sent to the actual robot for execution by using the \textit{Execute} button. An additional button for starting and stopping suction is located behind the robot arm with the label \textit{Start Suction}. This button both shows the suction state and acts as a toggle to switch the suction state. The recording is represented as a queue of robot states, which contains 3D end effector positions and suction states. The recording is sent to the robot system using two ROS services, the first one moves the robot arm to a given coordinate, and the second one activates or deactivates the suction. 

Fig. \ref{fig:vr_dec} shows the VR interface when the declarative control is activated. With this control, our goal is to enable the user to grab the cube objects and define a goal that the robot should achieve autonomously. In this case, cube objects become interactable, the user can reach for the cubes, grab them with the hand controller, and put them to a target position. After that, the user clicks on the \textit{Register Goal State} button for registering the desired goal state. Then, they can click on the \textit{Solve} button to make the robot find a plan that leads to the goal state. The plan is first executed by the digital twin of the robot, for showing it to the user. Similarly to procedural control, the user can use the \textit{Restart} button to start from scratch and use \textit{Execute} button to make the real robot execute the movements.

For planning, we use a PDDL solver. The initial states of the objects and the user-defined goal state makes up a PDDL problem. This problem and the PDDL domain are sent to an AI Planner (PDDL solver), that returns PDDL actions that lead from the initial state to the goal state. The PDDL domain that we use to describe our system is an extension of the famous Blocksworld \cite{blocksworld} domain. As shown in Listing \ref{listing:pddl_domain}, we extend this domain, by introducing a new type as \textit{position}. 

\begin{lstlisting}[label=listing:pddl_domain, caption=PDDL domain definition: types and predicates (extensions to original blocksworld domain are in red), escapechar=!]
(:types block !\color{purple}{position}!)
(:predicates
    (on ?x - block ?y - block)
    (ontable ?x - block)
    (clear ?x - block)
    (handempty)
    (holding ?x - block)
    !\color{purple}{(at ?x - block ?p - position)}!
    !\color{purple}{(free ?p - position)}!)
\end{lstlisting}

To describe possible states involving a position, we introduce two new predicates. Likewise, to describe actions involving a position, we introduce two new actions as \textit{place} and \textit{pick-from-pos}. Place action describes placing a block to a position, and pick-from-pos action describes picking up a block from a position (see Listing \ref{pddl_action_ext}).

\begin{lstlisting}[language=PDDL, label=pddl_action_ext, caption=PDDL actions of the extended blocksworld domain]
  (:action place
	     :parameters (?x - block ?p - position)
	     :precondition (and (holding ?x) (free ?p))
	     :effect
	     (and (not (holding ?x))
		   (not (free ?p))
		   (clear ?x)
		   (handempty)
		   (at ?x ?p)))
		   
  (:action pick-from-pos
	     :parameters (?x - block ?p - position)
	     :precondition (and (clear ?x) (at ?x ?p)
	     (handempty))
	     :effect
	     (and (not (at ?x ?p))
		   (not (clear ?x))
		   (not (handempty))
		   (free ?p)
		   (holding ?x)))
     
\end{lstlisting}

\begin{lstlisting}[label=listing:solution, caption=An example solution]
(pick-up cube_0)
(place cube_0 pos_0)
(pick-up cube_1)
(place cube_1 pos_1)
\end{lstlisting}

We use tags for identifying the cube objects and the positions in the scene. The problem description and the actions returned by the PDDL solver contains only the tags. Listing \ref{listing:solution}, shows an example solution, where the robot should \textit{pick-up} \textit{cube\_0}, \textit{place} it to \textit{pos\_0}, and \textit{pick-up} \textit{cube\_1}, \textit{place} it to \textit{pos\_1}. The exact end effector positions for picking up the cubes and placing them to positions are defined by finding the coordinates of the objects with the given tags.

\section{Evaluation}
\label{eval}

For evaluating our VR interface for human-in-the-loop adaptive systems, we conducted a usability evaluation with regard to effectiveness, efficiency, and user satisfaction. Based on these criteria, the main goal of the usability evaluation was to analyze and compare the advantages and disadvantages of both implemented human-in-the-loop strategies: procedural control and declarative control. 

\subsection{Evaluation Setup}
\label{evalSetup}

We conducted a usability study with 14 participants. None of the participants were advanced VR users, 7 of them never used a VR device before, 4 participants were beginners, and 3 participants were intermediate users. We asked them to intervene in the Dobot Magician system by using the VR interface with each of the control strategies to perform a task.

\begin{figure}[hbt!]
\centering
\begin{subfigure}{.30\textwidth}
  \centering
  \includegraphics[width=\linewidth]{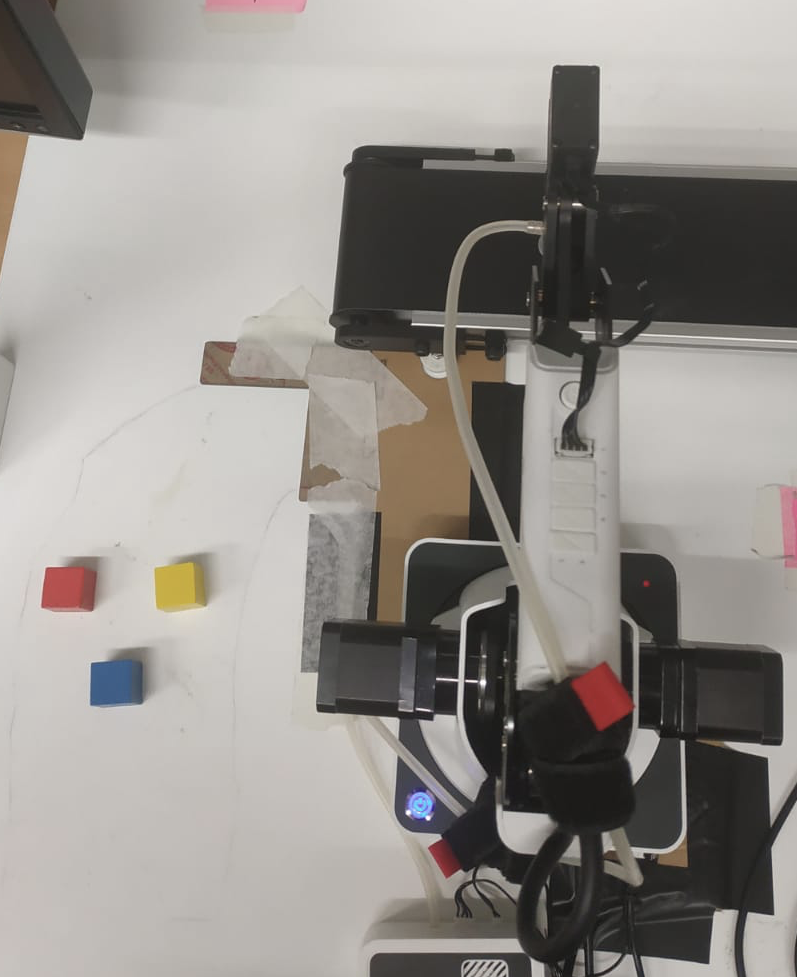}
  \caption{Initial state}
  \label{fig:task_before}
\end{subfigure}
\begin{subfigure}{.30\textwidth}
  \centering
  \includegraphics[width=\linewidth]{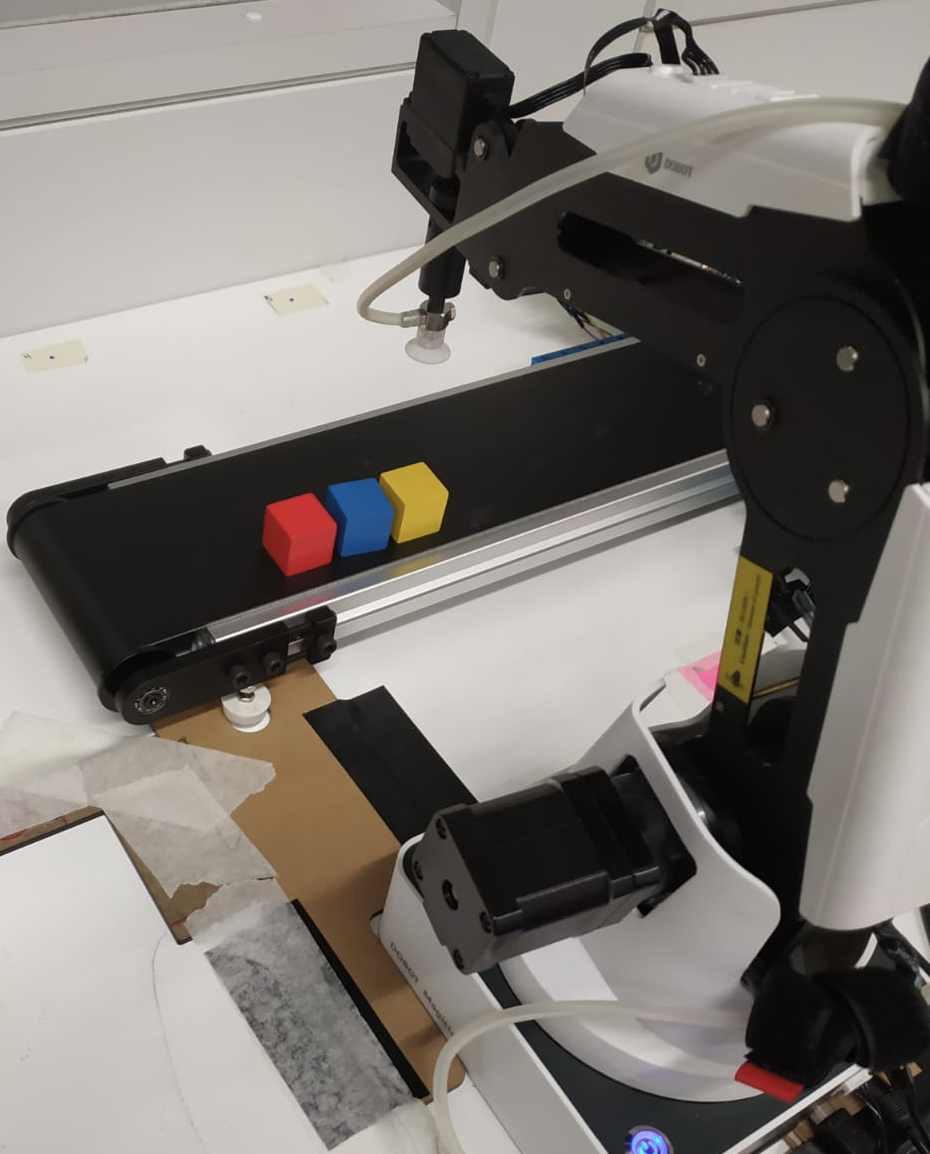}
  \caption{Goal state}
  \label{fig:task_after}
\end{subfigure}
\caption{\textit{Managed Element} before (a) and after task scenario completion (b)}
\label{fig:task_start_end}
\end{figure}

As shown in Fig. \ref{fig:task_start_end}, the task scenario consists of three cubes with different colors (red, blue, yellow) and three target positions also marked with the same three colors. The goal is to put each of the cubes to the target positions with the same color, i.e. to sort the cubes by their colors. Note that the target positions are only shown in the VR interface (Fig. \ref{fig:vr_procedural_vs_declarative}). To put a cube to a target position with the procedural control, the user needs to start recording, move the end effector of the arm on top of a cube, start suction, move the end effector to a target position, stop suction, stop recording, optionally replay recording, and execute. To do the same operation with declarative control, the user needs to grab a cube, put it to a target position, register the goal state, make the robot find a solution, and execute.

In each usability test session, we introduced the system components to the participants. Then, before the participant completed the task with each control strategy (procedural or declarative), we showed a video tutorial that demonstrates how that control strategy works in the VR interface. When the participant finished the tasks, we asked them to fill out a questionnaire and held an informal feedback interview to get further comments and hints for improvement.

\subsection{Efficiency}
\label{evalEfficiency}

For evaluating the system's efficiency, we measure task completion durations, by logging task start and end times. In procedural control, we consider clicking the record button as the start and clicking the execute button as the end of a session. In declarative control, we consider clicking the automated button (switching to declarative control) as the start and clicking the execute button as the end of a session. Finally, we calculate the efficiency for each control, as the average time taken to complete each test session.

\begin{table}[h!]
\centering
\begin{tabular}{c c c}
 & \multicolumn{1}{l}{Procedural Control} & \multicolumn{1}{l}{Declarative Control} \\
\textbf{Participant} & \multicolumn{1}{l}{\textbf{Completion Time (s)}} & \multicolumn{1}{l}{\textbf{Completion Time (s)}} \\
\hline
1 & 346.88 & 106.20 \\
2 & 301.82 & 101.45 \\
3 & 114.37 & 59.38 \\
4 & 52.34 & 84.71 \\
5 & 131.75 & 79.60 \\
6 & 87.14 & 56.25 \\
7 & 146.69 & 75.54 \\
8 & 94.39 & 69.35 \\
9 & 272.21 & 351.46 \\
10 & N/A & N/A \\
11 & 139.44 & 82.19 \\
12 & 70.02 & 167.56 \\
13 & 187.65 & 103.68 \\
14 & 298.47 & 116.14 \\
\hline
Average & 172.55 & 111.81\\
\hline
\end{tabular}
\caption{Task completion times for procedural and declarative control}
\label{completion_times}
\end{table}

As shown in Table \ref{completion_times}, the average task completion time in the procedural control strategy is 172.55 seconds and in the declarative control strategy, it is 111.81 seconds. The same task can be completed a little bit less than 3 minutes in procedural control, and less than 2 minutes in the declarative variant. Please note that no efficiency measures are available from the tenth participant as this one could place objects to correct positions but did not execute the program. In general, these efficiency results suit the user feedback from the interviews, where the participants have frequently noted that they were able to complete the task in declarative control strategy faster.

\subsection{Effectiveness}
\label{evalEffectiveness}

The effectiveness of the system was calculated for each of the control strategies as the number of successfully completed sub-tasks divided by the total number of sub-tasks. As described, the task scenario consists of placing three cubes of different colors to the positions that match the cubes’ colors. We consider moving each cube to their respective positions as a sub-task. Therefore, for three cubes we obtain three sub-tasks. We distinguish between the effectiveness of the VR interface and the managed element. Successful task completion in the VR interface does not indicate successful task completions by the managed element. The managed element's performance is affected by sensor data in the monitoring phase, e.g., the objects are not instantiated at their correct positions in VR. Therefore, we measure the effectiveness of the managed element separately.  

\begin{table}[h!]
\centering
\begin{tabular}{c|cc|cc}
{} &  \multicolumn{2}{c}{Procedural Control} & \multicolumn{2}{c}{Declarative Control}\\
\textbf{Part.} & \textbf{in VR} & \textbf{in Real} & \textbf{in VR} & \textbf{in Real} \\
\hline
1 & 3 & 2 & 3 & 3 \\
2 & 3 & 2 & 3 & 2 \\
3 & 3 & 3 & 3 & 2 \\
4 & 3 & 3 & 3 & 1 \\
5 & 3 & 3 & 3 & 3 \\
6 & 3 & 2 & 3 & 3 \\
7 & 3 & 2 & 3 & 3 \\
8 & 3 & 2 & 3 & 3 \\
9 & 3 & 3 & 3 & 2 \\
10 & 0 & 0 & 0 (3)* & 0 \\
11 & 3 & 2 & 3 & 3 \\
12 & 3 & 1 & 3 & 3 \\
13 & 3 & 3 & 3 & 3 \\
14 & 3 & 3 & 3 & 3 \\
\hline
\end{tabular}
\caption{Sub-task completion results for procedural and declarative control strategies in VR and the real physical world}
\label{effectiveness_records}
\end{table}

Table \ref{effectiveness_records} shows the sub-task completion results for each of the control strategies both for VR and for the real world. As mentioned before, the \textit{10th participant} could place the three cubes in correct positions but did not execute the program. Thus, we do not count this session as successful completion.

With these results, we calculate the effectiveness of the system as follows, where \(n=14\):

For procedural control strategy in VR, where \(T_{VR}(i)\) is the task completion count out of 3 for each participant:

\[
P_{VR} =
\sum\limits_{i=1}^n T_{VR}(i)\times {\dfrac{100}{n}}
\]

For procedural control strategy in the real world, where \(T_{RW}(i)\) is the task completion count out of 3 for each participant:

\[
P_{RW} =
\sum\limits_{i=1}^n T_{RW}(i)\times {\dfrac{100}{n}}
\]

For declarative control strategy in VR, where \(T_{VR}(i)\) is the task completion count out of 3 for each participant:

\[
D_{VR} =
\sum\limits_{i=1}^n T_{VR}(i)\times {\dfrac{100}{n}}
\]

For declarative control strategy in the real world, where \(T_{RW}(i)\) is the task completion count out of 3 for each participant:

\[
D_{RW} =
\sum\limits_{i=1}^n T_{RW}(i)\times {\dfrac{100}{n}}
\]

Therefore, we find effectiveness for each mode in VR and in the real world as follows:

\[P_{VR} = 92.86\%\] \[P_{RW} = 73.81\%\]
\[D_{VR} = 92.86\%\] \[D_{RW} = 80.95\%\]

With regard to effectiveness, we can see that there are no differences between the procedural and declarative control of the VR interface. Nearly all of the participants (despite one) could successfully finish the given task. The difference between the effectiveness of the VR interface and the managed element is expected, due to potentially imprecise sensor data. Here, the effectiveness of the managed element is 7\% better with declarative control. This is expected: while in declarative control the robot always picks up the cubes exactly from their centers, in procedural control, the user might not hit the exact center.

\subsection{User Satisfaction}
\label{evalSatisfaction}

For evaluating user satisfaction, firstly, we asked the participants to fill out the System Usability Scale (SUS) questionnaire \cite{sus_questionnaire}. It contains the following ten statements that are rated on a Likert scale between 1 (Strongly Disagree) and 5 (Strongly Agree): 

Q1: I think that I would like to use this system frequently.
Q2: I found the system unnecessarily complex.
Q3: I thought the system was easy to use.
Q4: I think that I need the support of a tech. person to use this system.
Q5: I found the various functions in this system were well integrated.
Q6: I thought there was too much inconsistency in this system.
Q7: I would imagine that most people would learn to use this system very quickly.
Q8: I found the system very cumbersome to use.
Q9: I felt very confident using the system.
Q10: I needed to learn a lot of things before I could get going with this system.

We asked the participants to fill out the SUS questionnaire separately for each control strategy. Table \ref{SUS_answers_proc} shows the results of the SUS questionnaire for the procedural control strategy and table \ref{SUS_answers_dec} for the declarative control strategy. Please note that answers are between 1 (Strongly Disagree) and 5 (Strongly Agree), while higher is better for items with odd numbers and lower is better for items with even numbers.

\begin{table}[]
\begin{tabular}{m{1em} m{1em} m{1em} m{1em} m{1em} m{1em} m{1em} m{1em} m{1em} m{1em} m{1em} m{3em}}
\hline
Par.    & Q1 & Q2 & Q3 & Q4 & Q5 & Q6 & Q7 & Q8 & Q9 & Q10 & Score \\
\hline
1       & 2   & 2   & 2   & 2   & 3   & 2   & 4   & 3   & 2   & 4    & 50      \\
2       & 2   & 4   & 4   & 1   & 5   & 4   & 4   & 2   & 4   & 1    & 67.5    \\
3       & 3   & 3   & 4   & 1   & 4   & 2   & 4   & 2   & 5   & 1    & 77.5    \\
4       & 2   & 4   & 3   & 2   & 2   & 4   & 3   & 4   & 2   & 4    & 35      \\
5       & 4   & 1   & 3   & 1   & 5   & 1   & 5   & 1   & 5   & 1    & 92.5    \\
6       & 2   & 4   & 3   & 2   & 4   & 3   & 4   & 2   & 4   & 2    & 60      \\
7       & 2   & 5   & 1   & 4   & 3   & 2   & 2   & 4   & 3   & 5    & 27.5 \\
8       & 3   & 2   & 4   & 3   & 4   & 1   & 4   & 2   & 3   & 1    & 72.5    \\
9       & 4   & 4   & 4   & 2   & 4   & 2   & 3   & 3   & 3   & 3    & 60      \\
10      & 2   & 2   & 2   & 4   & 3   & 3   & 3   & 2   & 2   & 5    & 40      \\
11      & 3   & 3   & 3   & 3   & 3   & 3   & 3   & 3   & 3   & 5    & 45      \\
12      & 3   & 2   & 4   & 2   & 4   & 1   & 4   & 2   & 4   & 2    & 75      \\
13      & 3   & 1   & 5   & 1   & 5   & 1   & 5   & 1   & 5   & 1    & 95      \\
14      & 4   & 1   & 4   & 1   & 5   & 1   & 5   & 1   & 5   & 1    & 95      \\
    \hline
Av. & 2,8 & 2,7 & 3,3 & 2,1 & 3,9 & 2,1 & 3,8 & 2,3 & 3,6 & 2,6  & 63.75  \\
    \hline
\end{tabular}
\caption{Results of the SUS questionnaire for the procedural control strategy}
\label{SUS_answers_proc}
\end{table}

\begin{table}[]
\begin{tabular}{m{1em} m{1em} m{1em} m{1em} m{1em} m{1em} m{1em} m{1em} m{1em} m{1em} m{1em} m{3em}}
\hline
Par.    & Q1 & Q2 & Q3 & Q4 & Q5 & Q6 & Q7 & Q8 & Q9 & Q10 & Score \\
\hline
1       & 5   & 1   & 5   & 2   & 4   & 1   & 5   & 1   & 4   & 2    & 77.5  \\
2       & 5   & 1   & 5   & 1   & 5   & 4   & 5   & 1   & 5   & 5    & 82.5  \\
3       & 3   & 2   & 4   & 1   & 4   & 2   & 4   & 2   & 5   & 1    & 80    \\
4       & 5   & 1   & 5   & 2   & 5   & 1   & 5   & 1   & 5   & 1    & 97.5  \\
5       & 3   & 1   & 5   & 1   & 5   & 1   & 4   & 1   & 5   & 1    & 92.5  \\
6       & 5   & 2   & 4   & 1   & 4   & 2   & 4   & 2   & 5   & 2    & 82.5  \\
7       & 5   & 1   & 5   & 2   & 4   & 2   & 5   & 1   & 5   & 3    & 87.5  \\
8       & 4   & 1   & 4   & 2   & 4   & 1   & 5   & 2   & 4   & 1    & 85    \\
9       & 4   & 2   & 4   & 2   & 4   & 2   & 3   & 3   & 4   & 2    & 70    \\
10      & 2   & 3   & 3   & 4   & 4   & 2   & 3   & 3   & 1   & 2    & 47.5  \\
11      & 4   & 3   & 4   & 4   & 5   & 2   & 4   & 3   & 3   & 4    & 60    \\
12      & 3   & 2   & 4   & 2   & 3   & 2   & 4   & 2   & 4   & 2    & 70    \\
13      & 3   & 1   & 5   & 1   & 5   & 1   & 5   & 1   & 5   & 1    & 95    \\
14      & 5   & 1   & 5   & 1   & 5   & 1   & 5   & 1   & 5   & 1    & 100   \\
    \hline
Av. & 4   & 1,6 & 4,4 & 1,9 & 4,4 & 1,7 & 4,4 & 1,7 & 4,3 & 2    & 80.54 \\
    \hline
\end{tabular}
\caption{Results of the SUS questionnaire for the declarative control strategy}
\label{SUS_answers_dec}
\end{table}

As a summary of the SUS questionnaire, we derived a SUS score of 63,75 for the procedural control strategy while the declarative control achieved a SUS score of 80,54. SUS scores can be interpreted by using the adjective rating scale \cite{sus_adjective_scale} for example. According to the adjective scale, procedural control (SUS=63.75) can be defined as between \textit{OK} and \textit{good} which is in acceptability range, and declarative control (SUS=80.54) can be defined as close to \textit{excellent} (where 25 corresponds to \textit{worst imaginable}, and 100 corresponds to \textit{best imaginable}).

As a second step, we asked the participants a subset of the presence questionnaire \cite{presence_questionnaire}, which is used for measuring user's feeling of being present in a virtual environment. We consider user's degree of presence experience as a criterion for their satisfaction. The factors that affect this experience are realism, the possibility to act, the possibility to examine, quality of the interface (low feedback delay, high frame rate), and haptic quality (appropriate use of haptic feedback). These aspects were covered by the following presence questions:  

PQ1: How natural were your interactions with the virtual environment?
PQ2: How much did the visual aspects of the environment involve you?
PQ3: How much did your experiences in the virtual environment seem consistent with your real-world experiences?
PQ4: Were you able to anticipate what would happen next in response to the actions that you performed?
PQ5: How well could you examine the objects?
PQ6: How much delay did you experience between your actions and expected outcomes?
PQ7: How well could you move or manipulate objects in the virtual environment?

As the way of providing presence does not change with different control strategies, when asking the presence questionnaire, we do not distinguish between the two control strategies and cover both of them.

\begin{table}[h!]
\centering
\begin{tabular}{m{1em} m{1em} m{1em} m{1em} m{1em} m{1em} m{3em} m{1em} m{1em} m{3em}}
\hline
Part. & PQ1 & PQ2 & PQ3 & PQ4 & PQ5 & PQ6(R) & PQ7 & Tot. & Pct.\\ 
 \hline
 1 & 6 & 6 & 4 & 4 & 6 & 2 (6) & 6 & 38 & 78\%\\
2 & 5 & 5 & 6 & 6 & 6 & 2 (6) & 6 & 40 & 82\%\\
3 & 5 & 5 & 6 & 5 & 6 & 5 (3) & 5 & 35 & 71\%\\
4 & 4 & 6 & 6 & 7 & 6 & 1 (7) & 6 & 42 & 86\%\\
5 & 6 & 1 & 6 & 6 & 6 & 2 (6) & 7 & 38 & 78\%\\
6 & 5 & 6 & 5 & 5 & 5 & 3 (5) & 6 & 37& 76\%\\
7 & 3 & 3 & 4 & 3 & 7 & 4 (4) & 7 & 31& 63\%\\
8 & 5 & 5 & 5 & 6 & 4 & 2 (6) & 5 & 36& 73\%\\
9 & 4 & 5 & 5 & 4 & 5 & 4 (4) & 5 & 32& 65\%\\
10 & 2 & 3 & 2 & 3 & 1 & 1 (7) & 1 & 19& 39\%\\
11 & 4 & 6 & 5 & 5 & 4 & 3 (5) & 5 & 34& 69\%\\
12 & 1 & 1 & 3 & 5 & 5 & 3 (5) & 4 & 24& 49\%\\
13 & 6 & 7 & 7 & 7 & 7 & 2 (6) & 7 & 47& 96\%\\
14 & 6 & 6 & 6 & 6 & 7 & 1 (7) & 7 & 45& 92\%\\
 \hline
Av. & 4.4 & 4.6 & 5 & 5.1	& 5.4 & 2.5(5.5)	& 5.5 & 35.6 & 73\% \\
 \hline
\end{tabular}
\caption{Participants' answers to the presence questionnaire (answers are between 1 (not at all) and 7 (completely), R is reversed score)}
\label{presence_answers_table}
\end{table}

Table \ref{presence_answers_table} shows the participants' answers to the presence questionnaire. Except for the PQ6, a higher answer is better for all of the questions. Therefore, for PQ6 we also present the reversed answers beside the original ones. For calculating the total score we summed each participant's answers to each question. For better comparability, we then converted these scores to percentage values. In general, we can observe that the user's feeling of presence is very good. 

Besides the questionnaire results above, we collected general feedback from the participant to better compare both control strategies to each other. Based on this feedback, referring to the procedural control, we could see that most of the participants liked the flexibility in taking the robot arm and controlling it in a natural and realistic way. With this regard, some of the participants also commented that they liked the controller vibration (haptic feedback when a robot collides with a cube). However, some of the participants also remarked that this type of control is complicated as it contains repetitive tasks (e.g., activating/deactivating the suction button).   
Concerning the declarative control strategy, most of the users liked the point that you do not have to define every step. With this regard they made statements such as "It is great to define only the goal state." or "I found this control more intuitive.". However, these users complained about the transparency of the robot system as the robot arm movements are calculated by the AI planner which was commented with expressions like "I disliked that it appears to work magically.". In general, it was interesting to observe that the trade-off between transparency (amount of feedback and context information) and control (manual vs. fully automated) is a complex design decision that is highly dependent on the SAS and its application domain. 

\subsection{Discussion}
\label{evalDiscussion}

Table \ref{table:eval} shows a summary of the overall evaluation results. 

\begin{table}[h!]
\begin{tabular}{@{}lll@{}}
\toprule
 & \begin{tabular}[c]{@{}l@{}}Procedural Control\end{tabular} & \begin{tabular}[c]{@{}l@{}}Declarative Control\end{tabular} \\ \midrule
Efficiency & 172 s & 112 s \\
\begin{tabular}[c]{@{}l@{}}Effectiveness of VR Interface\end{tabular} & 92.86\% & 92.86\% \\
\begin{tabular}[c]{@{}l@{}}Effectiveness of Man. Element\end{tabular} & 73.81\% & 80.95\% \\
\begin{tabular}[c]{@{}l@{}}Presence Questionnaire\end{tabular} & \begin{tabular}[c]{@{}l@{}}73\% 
\end{tabular} & \begin{tabular}[c]{@{}l@{}}73\% 
\end{tabular}\\
SUS & \begin{tabular}[c]{@{}l@{}}63.75 
\end{tabular} & \begin{tabular}[c]{@{}l@{}}80.54 
\end{tabular} \\
\bottomrule
\end{tabular}
\caption{Summary of the evaluation results}
\label{table:eval}
\end{table}

Efficiency results show the average task completion times (in seconds) with each control. It shows that declarative control provides 35.2\% faster task completion. This is an expected result because the number of interactions is less compared with declarative control. The effectiveness of the VR interface is the same with each control strategy. Differences can be seen regarding the effectiveness of the VR interface and the execution on the real Dobot Magician system. This is expected, due to potentially imprecise sensor data. In general, the user satisfaction data show promising results for the practical usage of the VR interface for human-in-the-loop involvement. 

Our results show that using a graphical representation, such as a VR interface, it is possible to provide transparency about the system state. This approach does not require users to have extensive domain knowledge to reason about the system. Moreover, with our approach, people who used the system for the first time, were able to intervene in the system to complete a task by involving in its self-adaptation loop. We report that with procedural control, the users need more time to complete the task, but they have more freedom to define what exactly the system should perform. On the other hand, with declarative control, the users save a significant amount of time to complete the task, gain a slight precision benefit due to automated motion calculation, but they have less freedom to express their intentions.

\subsection{Threats to Validity}
\label{evalThreats}

Most of the participants of the usability study had computer science or a similar technical background which might lead to a bias regarding the observed evaluation results. Moreover, the evaluation setting may have influenced the participants to choose a favorite control strategy in the VR interface. As we asked the participants to complete the same task once for each mode, we have noticed that some participants felt channeled to favor one of the control strategies and disfavor the other mode. This might have been eliminated by asking the participants to complete the questionnaire for the first strategy before testing the second one. Finally, even though we expect that the presented VR solution would be also beneficial for other SAS, the results need to be interpreted only in the context of robotics systems with the described setting.   

\section{Summary and Future Work}
\label{sum}

In this paper, we have discussed the benefit of human involvement in self-adaptive systems and the need for increased transparency and controllability in such human-in-the-loop adaptive systems. To address these issues, we have presented a solution approach for enhancing human-in-the-loop adaptive systems through digital twins and VR interfaces. For supporting the aspect of transparency, based on the concept of the digital twin, we represent a self-adaptive system and its respective context in a virtual environment. With the help of a virtual reality (VR) interface, we support an immersive and realistic human involvement in the self-adaptation loop by mirroring the physical entities of the real world to the VR interface. For supporting the aspect of controllability and integrating the human in the decision-making and adaptation process, we have implemented and analyzed two different human-in-the-loop strategies in VR: a procedural control where the human can control the decision making-process and adaptations through VR interactions (human-controlled) and a declarative control where the human specifies the goal state and the configuration is delegated to an AI planner (mixed-initiative). The solution approach which combines digital twin and VR interfaces has been evaluated regarding efficiency, effectiveness, and user satisfaction and shows great potential in increasing the transparency and control of human-in-the-loop adaptive systems. 

Still, there are some directions for future work. The developed system currently only supports the creation of cube blocks in VR. Depending on the application context of the underlying SAS, the objects in the real environment can differ and should be covered with the help of sophisticated object detection techniques. In the case of complex shapes, it is also possible to create and use CAD models of the required shapes. 

Furthermore, it could be beneficial to further improve the way of human involvement by enhancing the VR interface through multi-modal input and output (e.g., robot control over voice command or auditive feedback). Concerning human-involvement it is important to notice that our solution especially focuses on the decision-making phase of a specific type of human-in-the-loop SAS. In this context, further investigations are needed to apply our solution for different types of human-in-the-loop SAS (e.g., with more complex goals or distributed components).    

Finally, for improving the generality of our solution, we plan to apply and evaluate the presented approach in further application domains of self-adaptive systems to investigate its efficiency, effectiveness, and user satisfaction concerning other domains than robotics. 






\bibliographystyle{./bibliography/IEEEtran}
\bibliography{./bibliography/IEEEabrv,./bibliography/IEEEexample}

\end{document}